\documentclass[10pt,aps,prl,superscriptaddress]{revtex4-2}
\usepackage{natbib}
\usepackage{amsmath,amsfonts,amssymb,verbatim,mathtools}
\usepackage{amsfonts}
\usepackage{graphicx}
\usepackage[normalem]{ulem}
\usepackage{color}
\usepackage{dcolumn}
\usepackage{tabularx}
\usepackage{keyval}
\usepackage{multirow}
\usepackage[left]{lineno}

\newcommand{\QMI}{Quantum Matter Institute, University of British Columbia, Vancouver, British Columbia, Canada}

\newcommand{\UBC}{Department of Physics $\&$ Astronomy, University of British Columbia, Vancouver, British Columbia, Canada}

\newcommand{\Argonne}{Materials Science Division, Argonne National Laboratory, Lemont, IL, United States}

\newcommand{\MPI}{Max Planck Institute for Solid State Research, Heisenbergstraße 1, Stuttgart, Germany}

\newcommand{\CLS}{Canadian Light Source, Saskatoon, Saskatchewan, Canada}

\newcommand{\FIB}{Center for Nanoscale Materials, Argonne National Laboratory, Lemont, IL, United States}

\begin{document}

\title{Universal electronic structure of multi-layered nickelates 
\\via oxygen-centered planar orbitals}

\author{Christine C. Au-Yeung}
\thanks{These authors contributed equally}
\affiliation{\QMI}
\affiliation{\UBC}

\author{X. Chen}
\thanks{These authors contributed equally}
\affiliation{\Argonne}

\author{S. Smit}
\thanks{These authors contributed equally}
\affiliation{\QMI}
\affiliation{\UBC}

\author{M. Bluschke}
\thanks{These authors contributed equally}
\affiliation{\QMI}
\affiliation{\UBC}

\author{V. Zimmermann}
\affiliation{\QMI}
\affiliation{\UBC}
\affiliation{\MPI}

\author{M. Michiardi}
\affiliation{\QMI}
\affiliation{\UBC}

\author{P.~C.~Moen}
\affiliation{\QMI}
\affiliation{\UBC}

\author{J. Kraan}
\affiliation{\QMI}
\affiliation{\UBC}

\author{C. S. B. Pang}
\affiliation{\QMI}
\affiliation{\UBC}

\author{C. T. Suen}
\affiliation{\QMI}
\affiliation{\UBC}
\affiliation{\MPI}

\author{S. Zhdanovich}
\affiliation{\QMI}
\affiliation{\UBC}

\author{M. Zonno}
\affiliation{\CLS}

\author{S. Gorovikov}
\affiliation{\CLS}

\author{Y. Liu}
\affiliation{\FIB}

\author{G. Levy}
\affiliation{\QMI}
\affiliation{\UBC}

\author{I. S. Elfimov}
\affiliation{\QMI}
\affiliation{\UBC}

\author{M. Berciu}
\affiliation{\QMI}
\affiliation{\UBC}

\author{G. A. Sawatzky}
\affiliation{\QMI}
\affiliation{\UBC}

\author{J. F. Mitchell}\email{ayching@phas.ubc.ca; mitchell@anl.gov; damascelli@physics.ubc.ca}
\affiliation{\Argonne}

\author{A. Damascelli}
\email{ayching@phas.ubc.ca; mitchell@anl.gov; damascelli@physics.ubc.ca}
\affiliation{\QMI}
\affiliation{\UBC}

\begin{abstract} 

{\bf In a series of groundbreaking discoveries, superconductivity has been demonstrated in the family of  multi-layered nickelates La$_3$Ni$_2$O$_7$ and La$_4$Ni$_3$O$_{10}$, with $T_c$ as high as 91 and 30K respectively under moderate pressure. Key questions remain open regarding the low-energy electronic states that support superconductivity in these compounds. Here we take advantage of the natural polymorphism between bilayer (2222) and alternating monolayer-trilayer (1313) stacking sequences that arises in bulk La$_3$Ni$_2$O$_7$ crystals, and by employing angle-resolved photoemission spectroscopy (ARPES) we identify a universal low-energy electronic structure in this family of materials. We observe the fingerprint of a doping-dependent spin-density wave (SDW) instability -- strong and coherent enough to reconstruct the Fermi surface, both by gapping out regions of the low-energy electronic structure as well as translating the $\beta$ pocket by a vector $Q_{t\beta}$ consistent with the results of previous neutron and x-ray scattering experiments. This demonstrates a universal connection between magnetism and fermiology in these materials. Using an effective tight-binding model, we simulate the spectral weight distribution observed in our ARPES dichroism experiments and establish that the low-energy electronic phenomenology is dominated by oxygen-centered planar orbitals, which -- upon moving along the Fermi surface away from the Ni-O-Ni bond directions -- evolve from the $d_{3x^2-r^2}$ and $d_{3y^2-r^2}$ symmetry characteristic of 3-spin polarons (3SP) to the familiar $d_{x^2-y^2}$ Zhang-Rice singlets (ZRS) that support high-temperature superconductivity in cuprates. By inclusion of magnetic moments on plaquettes of oxygen orbitals in our model, we show that ZRS-like states mediate the SDW. Combined with the observation that oxygen annealing is required to induce superconductivity in both thin films and bulk La$_3$Ni$_2$O$_7$, this demonstrates that the ZRS population dictates whether the ground state favors density-wave order or superconductivity – with hole doping suppressing the former and stabilizing the latter, as in the cuprates. Despite the multiorbital nature of the nickelates, our work thus establishes an empirical correspondence between their low-energy electronic structure and that of the cuprates, suggesting a common origin for unconventional superconductivity in both families.
}
\end{abstract}

\maketitle

\newpage
\subsection{Introduction}

Although unconventional superconductivity in the cuprates was discovered nearly four decades ago~\cite{ Bednorz1986}, the underlying mechanism responsible for the high-temperature Cooper pairing observed in this class of materials remains the subject of intense research today.  Since their discovery, researchers have explored materials with structural and electronic similarities to the cuprates~\cite{Maeno1994, Yan2015, Kim2015}, aiming to uncover deeper insights into high-$T_\text{c}$  superconductivity. The family of rare-earth nickelates represents a particularly successful example of this line of investigation, beginning with theoretical predictions of cuprate-analogous electronic structures~\cite{Anisimov1999, Chaloupka2008, Hansmann2009}. However, where the cuprates have only a single half-filled orbital of $d_{x^2-y^2}$ symmetry making up the lowest lying electronic states, the nickelates should possess two partially filled orbitals, namely the $d_{3z^2-r^2}$ and $d_{x^2-y^2}$. This extra degree of freedom raises intriguing questions regarding the similarities and differences in the underlying physics driving superconductivity in the two systems. Efforts to engineer superconductivity in the nickelates achieved initial success with the discovery of superconductivity in the infinite layer Nd$_{0.8}$Sr$_{0.2}$NiO$_2$ near 20~K~\cite{Li2019}. More recently, superconductivity was observed in bulk La$_3$Ni$_2$O$_7$ (LNO327) and trilayer La$_4$Ni$_3$O$_{10}$ (LNO4310) under high pressure up to 91~K~\cite{Sun2023,Li2025,zhu2024}, as well as in LNO327 thin films at ambient pressure under compressive epitaxial strain~\cite{Ko2024,zhou2024}. LNO327 is remarkable not only for its superconductivity but also for its intriguing crystal structure. For decades thought to be stable only in a bilayer (2222) Ruddlesden-Popper (RP) phase stacking, a stoichiometrically equivalent material consisting of alternating monolayer and trilayer stacking (1313) has recently been observed~\cite{Chen2023,Puphal2024}. This novel form of long-range polymorphism is exceptionally unusual and represents a first in this group of materials. To date, it is still unclear whether both phases contribute equally to superconductivity. 

The polymorphism poses a challenge for studying the electronic structures of these intertwined phases using bulk-sensitive techniques such as DC electrical transport. At the same time, it opens up valuable opportunities to systematically investigate the impact of these distinct structural motifs on the electronic states most relevant to superconductivity. To do so, we employ angle-resolved photoemission spectroscopy (ARPES), a direct probe of the momentum-resolved electronic structure. In 2024, several groups reported ARPES measurements of LNO327; however, the crystals studied either had unspecified structural compositions \cite{Yang2023}, or exclusively featured only one structure omitting the other \cite{Abadi2024}. Here we report, to the best of our knowledge, the first comparative ARPES investigation of pure 2222- and 1313-LNO327 single crystals. Our ARPES results demonstrate that, while the 2222 and 1313 phases of LNO327 are discernible based on their valence band spectra, the key features of the low energy electronic structure are strikingly similar, and are in fact also shared by LNO4310~\cite{du2024,li2017}. In addition, ARPES dichroism reveals the oxygen-centered planar orbital nature and momentum-dependent symmetry of the first electron removal states of LNO327. As we move away from the Ni–O–Ni bond directions along the Fermi surface (FS), we observe the evolution from the $d_{3x^2-r^2}$ and $d_{3y^2-r^2}$ symmetry -- characteristic of bond-centered 3-spin polarons (3SP), to the $d_{x^2-y^2}$ symmetry -- akin to the familiar Zhang-Rice singlets (ZRS) in cuprates. Within the appropriate effective model, the observed phenomenology suggests that the latter symmetry take a more prominent role in the low-energy electronic structure of the nickelates, highlighting their close connection to the incommensurate spin-density wave (SDW) and potentially also superconductivity. For clarity, throughout this work we refer to the first electron removal states with $d_{x^2-y^2}$ symmetry as ZRS-like states, and those with $d_{3x^2-r^2}$/$d_{3y^2-r^2}$ symmetry as 3SP-like states, even though in our effective tight-binding model, their corresponding spin correlations are ignored [see Sect.\,S11 in Supplementary Information (SI) for a more detailed overview].

\begin{figure}[t!]
\includegraphics[width=16cm]{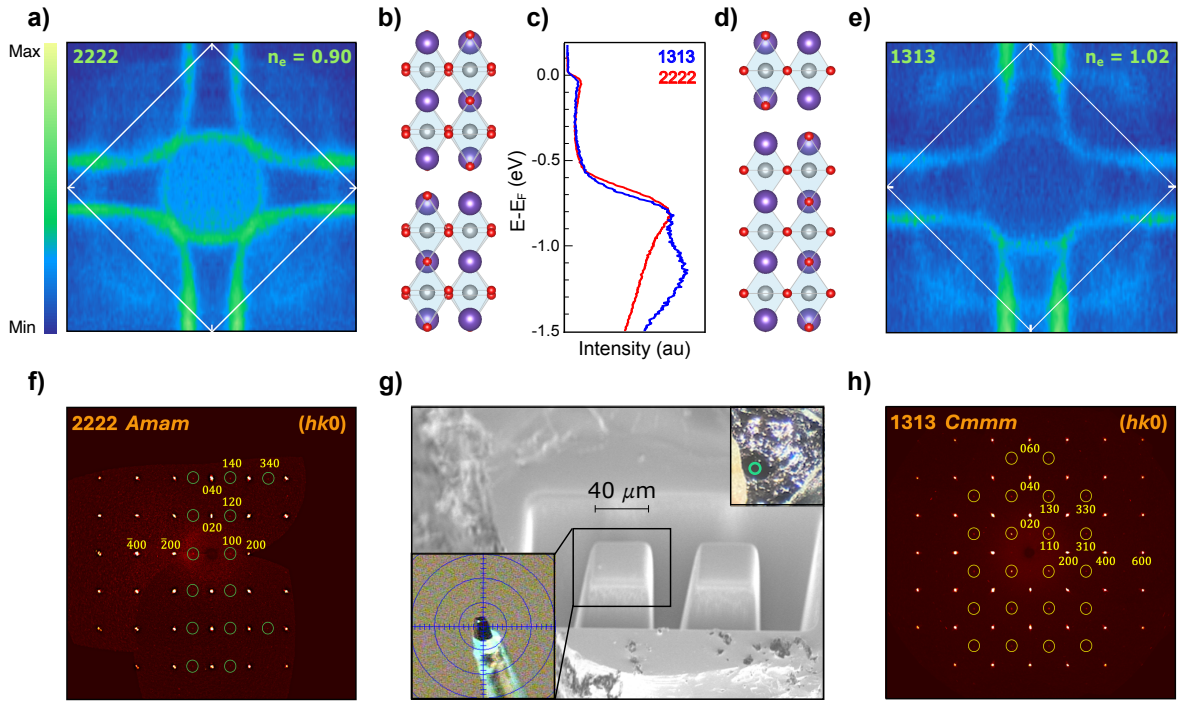}
\caption{{\bf Electronic structure of polymorph LNO327.}  \textbf{(a,e)} FS of 2222-LNO327 and 1313-LNO327 measured with 100~eV photons, integrated over light polarization and symmetrized around $k_x = 0$ of the tetragonal BZ  (white diamond indicates the orthorhombic BZ), and \textbf{(b,d)} corresponding lattice structure (blue: oxygen; red: lanthanum; grey: nickel). The number of electrons $n_e$ is calculated as the ratio between the area of the $\alpha$ and the $\beta$ FS as labeled (counting electrons and including spin degeneracy), and the area of the tetragonal BZ (see SI Sect.\,S5). \textbf{(c)} Valence band ARPES spectra integrated at the tetragonal zone boundary for both the 2222 and 1313 phases. \textbf{(f,h)} Integrated ($h$~$k$~0) zone XRD patterns of extracted crystallites on which the FS in (a,e) were taken, with diffraction spots showing the pure 2222 and pure 1313 structure, respectively. \textbf{(g)} Extraction of crystallites via ion milling: top inset, image of cleaved LNO327 after ARPES experiments, with green circle indicating the region measured by ARPES; bottom inset, extracted section of crystal for XRD of approximately 40 × 40 $\mu\text{m}^2$ wide and 20~$\mu\text{m}$ thick.}
\label{Fig1}
\end{figure}
 
\subsection{Fermiology and valence bands of 1313/2222-LNO327 }
We have performed ARPES measurements on LNO327 crystals of approximately 1×1 mm$^2$ size, known from x-ray diffraction to contain a mixture of the 1313 and 2222 phases. Careful spatial mapping using photoemission revealed two distinct regions present in all crystals measured. These regions are identifiable by their characteristic valence band spectra, particularly at binding energies $E_{\text{B}}~>~0.5$~eV [Fig.\,\ref{Fig1}(c) and Sect.\,S1 and S2 in SI]. The typical domain size of these regions was found to be $\sim 500\times500~\mu\text{m}^2$, and post-ARPES x-ray diffraction on the same exact crystallites -- extracted by ion milling [Fig.\,\ref{Fig1}(g) and Fig.\,S1] -- confirmed these regions with their distinct spectral signatures to be the phase pure 2222 and 1313 structures [Fig.\,\ref{Fig1}(f), (h)]. Surprisingly, the low-energy electronic structure of the 2222 and 1313 phases are nearly identical, as is evident from the FS shown in Fig.\,\ref{Fig1}(a) and \ref{Fig1}(e). Both FS consist of a central circular electron pocket [labeled as the $\alpha$ FS] and large square hole pockets [the $\beta$ FS] around the corners of the tetragonal Brillouin zone (BZ), corresponding to approximately 10$\%$ and 60$\%$ of the BZ volume, respectively. The number of electrons $n_e$ is then calculated as the ratio between the area of the combined FS -- counting electrons and multiplied by two assuming spin degeneracy -- and the area of the tetragonal BZ for simplicity (SI Sect.\,S5). The actual ambient-pressure lattice structure of LNO327, however, is orthorhombic, arising from an in-plane $\sqrt{2}\times\sqrt{2}$ distortion that doubles the tetragonal unit cell and rotates it by 45$^\circ$ in real space, thereby folding the $\beta$ FS into the $\beta'$ FS in reciprocal space [Fig.\,\ref{Fig1}(a,e), see also SI Sect.\,S13]. Besides the Luttinger volume changing from $n_e\!=\!0.9$ (2222-LNO327) to $1.1$ (1313-LNO327), the fermiology of the two phases is very similar and much simpler than predicted by density functional theory (DFT) \cite{Zhang2024,LaBollita2024}. It is also consistent with published data on LNO327 as well as trilayer LNO4310~\cite{Yang2023,du2024,Li_2024}, with one exception that reported instead a much more complex and DFT-like fermiology for 1313-LNO327 \cite{Abadi2024} (SI Sect.\,S5 and S9). This is reminiscent of the dichotomy between under- and overdoped cuprates, where bilayer- and trilayer-split bonding and antibonding FS are observed only beyond optimal doping with the weakening of electron-electron correlations \cite{Fournier2010,Ding1996,Feng2001,Chuang2001,Luo2023}, and progressively merge into one another upon underdoping as correlations increase. Thus, the apparent discrepancy in the reported fermiology for nickelates points to a varying degree of electron filling and in turn many-body correlations, potentially associated with different oxygen contents \cite{Abadi2024,Zhou2025,Wang_2025}.

\begin{figure}[t!]
\includegraphics[width=16cm]{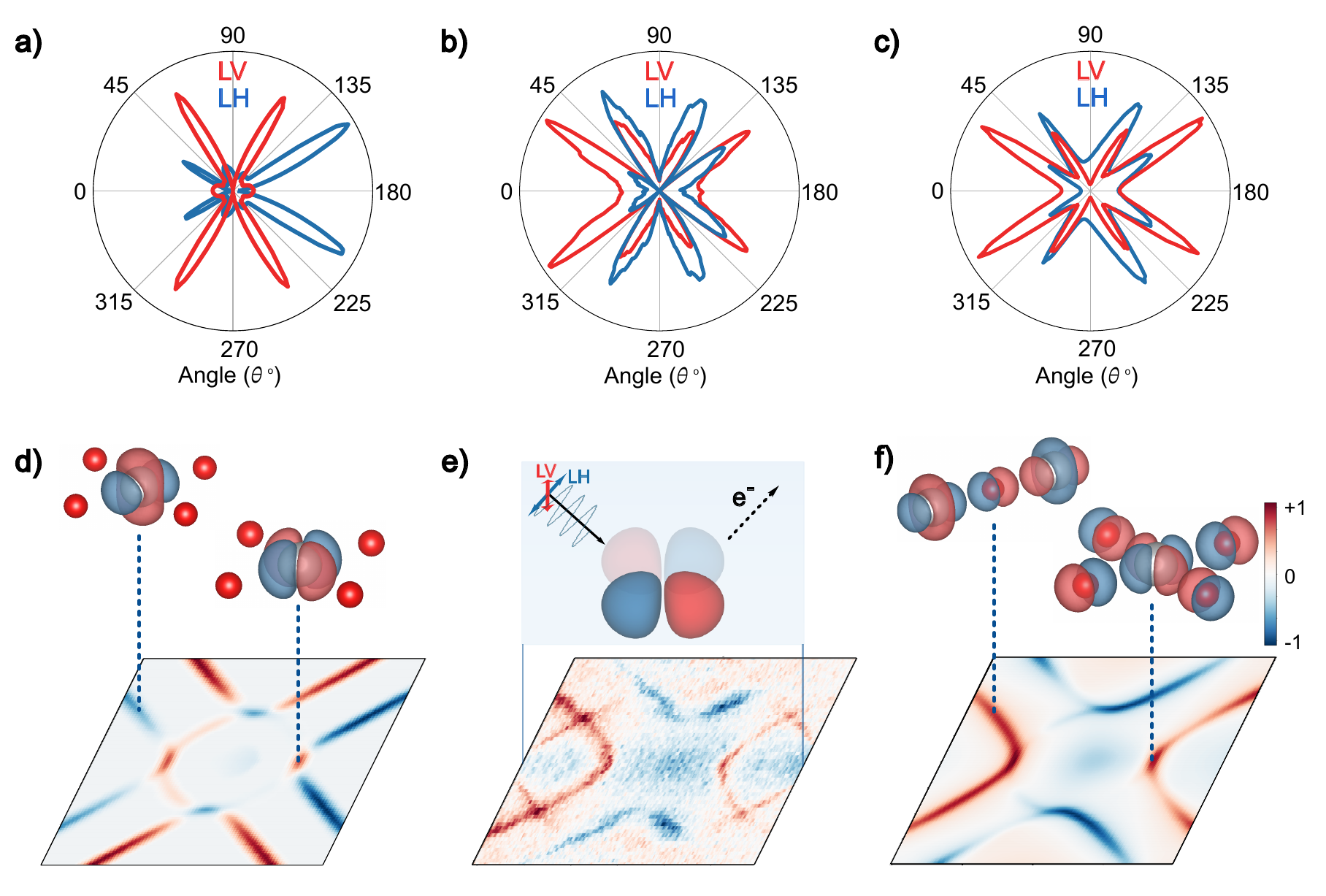}
\caption{{\bf Linear dichroism and symmetry of electronic wavefunctions.} \textbf{(a-c)} Polar plot of the simulated LNO327 ARPES intensity for linear vertical and linear horizontal (LV, LH) polarizations in the first BZ, where the distance from the center defines the magnitude of the signal at a given angle along the FS. The polar plots are obtained from \textbf{(a)} an effective tight-binding model with only Ni $e_g$ orbitals, \textbf{(b)} experiment, and \textbf{(c)} an effective tight-binding model with both Ni $e_g$ and O~$p$ orbitals. \textbf{(d,f)} (top) Schematic representation of wavefunctions at the indicated momenta along the FS with relative phases in red and blue (O and Ni sites in red and silver, respectively), along with (bottom) the simulated ARPES dichroism (LV-LH) from the same model as in \textbf{(a,c)}. \textbf{(e)} (top) Experimental geometry, showing the polarization vector of the light with respect to the in-plane orbitals, along with (bottom) the experimental dichroism measured by ARPES on 1313-LNO327.}
\label{Fig2}
\end{figure}

\subsection{Fermi surface orbital decomposition from ARPES dichroism}

The similarity in the Fermi surfaces of the two polymorph phases, despite the differences in the valence bands, suggests that both are susceptible to the same low-energy phenomena. In order to reveal the orbitals and symmetries relevant to the low-energy electronic structure, we have conducted a comprehensive analysis of the polarization-dependent matrix elements that modulate the measured ARPES intensity; while this is discussed here based on  data from 1313-LNO327 (Fig.\,\ref{Fig2}), an equivalent phenomenology is observed for 2222-LNO327 (SI Sect.\,S3). In our experimental geometry, as shown in Fig.\,\ref{Fig2}(e) (top), the measurement plane is oriented at 45$^{\circ}$  to the Ni-O-Ni bond. Linear vertical (LV) polarization corresponds to a polarization vector parallel to the sample surface with an odd symmetry with respect to the measurement plane. The polarization vector in linear horizontal (LH) is instead oriented at 45$^{\circ}$ with respect to the sample surface and has an even symmetry with respect to the measurement plane. Fig.\,\ref{Fig2}(e) (bottom) presents the linear dichroism (LV~$-$~LH) at the Fermi energy ($E_{\text{F}}$), showing the intensity difference generated by the matrix elements. The intensity of the $\alpha$ band is similarly weak in both LV and LH polarization at $E_{\text{F}}$, leading to a lack of dichroic signal, while most of the dichroism originates from the $\beta$ band. LV and LH strongly modulate the intensity between different branches of the $\beta$ FS; however, within the same branch, the relative intensity between LV and LH remains roughly constant. For a more quantitative polarization comparison, we plot the ARPES intensity along the FS in polar coordinates in Fig.\,\ref{Fig2}(b). The $r$-axis represents the integrated intensity along the radial direction of the first BZ at the indicated angle, where the $\Gamma$ point defines the center of rotation and the measurement plane defines the $\theta = 0^{\circ}$ direction (see also SI Sect.\,S4). The strong intensity variation between LV (red) and LH (blue) offers a valuable opportunity to study the symmetry of the first electron removal state along the FS, by using the \textit{chinook} package \cite{Day2019} to simulate the ARPES intensity with tight-binding models that correctly reproduce the shape of the experimental FS.

To this end, the challenge is to develop a minimal tight-binding model (see also SI Sect.\,S6-S8) that agrees with the simple two-pocket FS observed here by ARPES  on both 2222- and 1313-LNO327, and more generally on LNO327 as well as LNO4310 \cite{Yang2023,du2024,li2017}, while DFT predicts a much more complex fermiology for all these compounds \cite{Puphal2024,Chen2023,li2017,du2024}. To reconcile this discrepancy we note that, at some electron fillings, strong electron correlations may gap out certain bands. In the case of 2222-LNO327, this is achieved in DFT calculations with the inclusion of the effective on-site Coulomb interaction $U$ \cite{Yang2023}, and the minimal model naturally requires only a single NiO$_2$ bilayer (SI Sect.\,S6 and S7). Similarly, for 1313-LNO327 systematic theoretical studies based on DFT/DMFT consistently predict self-doping (Fig.\,S9) between the different planes \cite{Lechermann2024,LaBollita2024,Zhang2024}, which may result in a \textit{layer-dependent Mott transition} (SI Sect.\,S8) to correlated insulating states associated with a Ni $d^8$ occupation on the 1313 monolayer \cite{LaBollita2024} (2~$e_g$ electrons, $S\!=\!1$) as in NiO \cite{Zaanen1985}, and a Ni $d^7$ occupation on the trilayer inner plane (1~$e_g$ electron, $S\!=\!1/2$) akin to the electronic structure of insulating NdNiO$_3$ \cite{Johnston2014}. With the monolayer being fully gapped and only weakly coupled to the trilayer, these correlation effects can be captured in a tight-binding trilayer model for a suitable choice of parameters that push the corresponding inner plane bands away from $E_{\text{F}}$, resulting once again in only two FS pockets as in the minimal bilayer model (SI Sect.\,S8). Thus, the bilayer model appears to be sufficient to describe  the basic fermiology and low-energy band structure observed in ARPES for both 1313 and 2222 polymorphs of LNO327, as well as trilayer LNO4310; as such, it might be expected to contain the essential characteristics of the first electron removal state in the whole family of layered nickelates: the $\alpha$ ($\beta$) FS being a symmetrical (anti-symmetrical) linear combination of the coupled layers via the Ni $d_{3z^2-r^2}$ and the interplanar O $p_z$ orbitals, except at their symmetry-protected degeneracy along the vertical plane at 45$^\circ$ with respect to the Ni-O-Ni bond direction, where mirror symmetry forbids hybridization between the Ni $d_{x^2-y^2}$ (odd) and $d_{3z^2-r^2}$ (even) orbitals hence decoupling the layers.


We then start with constructing an effective bilayer tight-binding model using only Ni $e_g$ orbitals, with the contribution of all oxygen states projected out (SI Sect.\,S6). However, the linear dichroic ARPES intensity simulated using this band structure in \textit{chinook} \cite{Day2019} [Fig.\,\ref{Fig2}(a,d)] does not at all match the experimentally observed dichroism presented in Fig.\,\ref{Fig2}(b,e). On the contrary, we show that the inclusion of the neighboring in-plane oxygen $p_x$ or $p_y$ orbitals -- in addition to the Ni $e_g$ orbitals -- is necessary to successfully reproduce the experimental dichroic intensity for 1313-LNO327 [Fig.\,\ref{Fig2}(c,f)] as well as 2222-LNO327 (SI Sect.\,S7). Comparing the resulting wavefunctions in Fig.\,\ref{Fig2}(d) and (f), the most noticeable differences are the prominent in-plane character and the larger spatial extent of the wavefunctions in Fig.\,\ref{Fig2}(f), which involve the neighboring oxygen sites, highlighting the significant hole density present on the Ni-O-Ni bond. This establishes the importance of the strong Ni and O hybridization, resulting in more extended, molecular-like rather than atomic in-plane orbitals. Along the direction at 45$^{\circ}$ with respect to the Ni-O-Ni bonds, the electronic states consist of linear combinations of Ni $d_{x^2-y^2}$ and planar O $p_{x,y}$ orbitals, reminiscent of the spatial orbital character of the ZRS realized in the vicinity of $(\pi/2,\pi/2)$ along the superconducting cuprate nodal direction \cite{Zhang1988,Lau2011}. 
By contrast, along the two perpendicular Ni-O-Ni bond directions (in the vicinity of the antinodal points in the cuprate analogy), the electronic wavefunctions are described by linear combinations of planar O $p_{x}$ and Ni $d_{3x^2-r^2}$, or O $p_{y}$ and Ni $d_{3y^2-r^2}$ orbitals, respectively. These states are reminiscent of the spatial orbital character of the 3SP proposed in cuprates~\cite{Emery1988} (see further discussion in SI Sect.\,S11). 


\subsection{Fermi surface reconstruction from electronic order}
 Common to multi-layer nickelates, in addition to the expected $\alpha$  and $\beta$ bands, an anomalous third band [labeled here $t\beta$ for `translated beta', see Fig.\,\ref{Fig3}(a)] has been consistently observed at the orthorhombic zone boundary in both LNO327 and LNO4310 using 75~eV photons \cite{Yang2023,Abadi2024, du2024,li2017}. We observe the same $t\beta$-derived FS at 75~eV and 45~eV but, interestingly, the strong intensity modulation with photon energy causes it to be completely absent at 100 eV [see Fig.\,\ref{Fig1}(a,e)] and 7 eV \cite{Yang2023,du2024}. This feature has been described in the literature as either due to an impurity phase or as an anomalous surface state \cite{Yang2023, Abadi2024}. However, the persistent appearance of the $t\beta$ band suggests the possibility that it is a universal feature in this family of materials. Moreover, this band is unusual because it does not seem to form a closed FS pocket, in contrast to the $\alpha$ and $\beta$ bands; instead of being one continuous FS, it appears as fragments of two quasi-1D FS, perpendicular to one another and crossing along the $k_x = 0$ plane. 

\begin{figure}[t!]
\includegraphics[width=17cm]{Figures/Fig3_v2.png}
\caption{{\bf Anomalous electron pocket originating from doping-dependent incommensurate band folding.} \textbf{(a)} ARPES spectra and corresponding FS from 1313-LNO327, measured with 45~eV photons and symmetrized around $k_y = 0$. Cut 1 is taken at $k_x\!=\!-0.61$, cut 2 at $k_x\!=\!0$, cut 3 at $k_x\!=\!0.61$, and cut 4 at $k_y\!=\!-0.5$ (expressed in reciprocal lattice units of the orthorhombic BZ); orange arrows indicate the momentum transfer $Q_{t\beta}$ connecting the $\beta$ band to the translated $t\beta$ band. Inset: the gray solid line shows the orthorhombic BZ, with its high-symmetry points labeled. 
The $\beta$ FS is highlighted with continuous yellow lines; it is translated by the orange arrow ($Q_{t\beta}$) to the $t\beta$ FS, with continuous and dashed segments highlighting portions detected and undetected by ARPES, respectively. \textbf{(b)} Left inset: Tight-binding FS translation by $q_1$, which maximizes the overlap of the circular portions of the $\alpha$ and the $\beta$ FS; right inset: translation by $q_2$, which maximizes the overlap of the straight portions of the $\beta$ FS. Main panel: Tight-binding FS autocorrelation along the $k_x$ direction as a function of the Luttinger volume $n_e\!=\![0.84,1.21]$; the dashed orange and green lines follow the $n_e$-dependence of $q_1$ and $q_2$, respectively (expressed in reciprocal lattice units of the orthorhombic BZ). The autocorrelation calculations for the full FS with $n_e\!=\!0.84$ and the ZRS-projected FS are highlighted with light blue gradient, and correspond to the same effective electron counting [see SI Sect.\,S12 and Fig.\,S13(b)]. \textbf{(c)} Comparison of simulated ($q_{1,2}$) and experimental ($Q_{t\beta}$) translation vectors versus the number of electrons $n_e$, with ARPES-derived data from this work (2222-LNO327 and 1313-LNO327), as well as from the literature for LNO327 (Yang et al. \cite{Yang2023}) and LNO4310 (Li et al. \cite{li2017}; Du et al.\cite{du2024}).}
\label{Fig3}
\end{figure}

In the following, we will focus the discussion once again on data from 1313-LNO327; note however that equivalent results and phenomenology are observed for 2222-LNO327 (Figs.\,S5 and S12). In Fig.\,\ref{Fig3}(a), we compare the dispersion of the $t\beta$  and $\beta$ bands. Cuts 1 and 3 are taken at $k_x = -0.61$ and $+0.61$, respectively, in reciprocal lattice units of the orthorhombic BZ and show the dispersion of the original $\beta$ band, while cut 2 shows the dispersion of the $t\beta$ band at $k_x = 0$; all three bands have the same velocity of $1.3\!\pm 0.5$ eV\AA. Cut 4 shows the dispersion along $k_y = -0.5$, with the $\beta$ and $t\beta$ bands once again highlighted, also having identical velocities of $0.5\!\pm 0.2$~eV\AA. To summarize, the visible part of the $t\beta$ FS appears as a translation of the original $\beta$ FS by $Q_{t\beta} = \pm(0.61, 0)$, as highlighted by the orange arrows. This translation vector indicates an \textit{incommensurate} band folding at 45$^{\circ}$ with respect to the Ni-O-Ni bond direction, which is inconsistent with the folding expected from common octahedral rotations in perovskites \cite{Glazer1972} (SI Sect.\,S13), strongly suggesting a symmetry breaking of electronic rather than structural origin.

Suggestively, the portion of the $t\beta$ FS that is not directly visible in the data is expected to be overlapping with the $\alpha$ FS near the zone center. Inspired by this visually apparent nesting, we compute the autocorrelation function along $k_x$ of our effective tight-binding model, which reproduces the experimental unfolded FS within a small range of doping. These simulations, shown in the insets of Fig.\,\ref{Fig3}(b), suggest that there are two doping-dependent wavevectors connecting significant portions of the unreconstructed FS: a longer wavevector $\pm$($q_1$, 0) connecting the curved portion of $\alpha$ and $\beta$ FS [Fig.\,\ref{Fig3}(b) left inset], as well as a shorter wavevector $\pm$($q_2$, 0) originating from the straight portions of the $\beta$ FS [Fig.\,\ref{Fig3}(b) right inset], both exhibiting a monotonic and approximately linear increase with Luttinger volume [Fig.\,\ref{Fig3}(b)]. Since the magnitude of the translation vector depends sensitively on the shape and volume of the FS, establishing its relationship with $Q_{t\beta}$ requires a quantitative comparison. To this end, we extracted the Fermi volume and $Q_{t\beta}$ from all our experimental data for both the 1313- and 2222-LNO327 polymorphs, as well as from published FS data for LNO327 \cite{Yang2023} and LNO4310 \cite{li2017,du2024} (Fig.\,S5). The results show remarkable agreement between our simulation of the doping-dependent $q_1$ translation vectors and the experimental $Q_{t\beta}$ values [Fig.\,\ref{Fig3}(c)]. Notably, translating the FS with the $q_1$ vector brings the ZRS region of the FS [Fig.\,\ref{Fig2}(f)] into overlap and qualitatively reproduces the visible part of the anomalous $t\beta$ FS observed by ARPES, whereas translating the FS by $q_2$ overlaps the 3SP region and shows no resemblance to the experimental FS. We therefore recompute the autocorrelation only for states corresponding to linear combinations of oxygen orbitals with ZRS symmetry (SI Sect.\,S12), which yields exclusively $\pm(q_1,0)$, as shown in the bottom curve of Fig.\,\ref{Fig3}(b). Together with its incommensurability and continuous evolution with doping, these results demonstrate the electronic origin of the band folding that gives rise to the $t\beta$ band via the ZRS-like states.

LNO327 and LNO4310 have been shown to be very similar compounds. Not only do both exhibit superconductivity under pressure and are characterized by strikingly similar FS including the consistent observation of the $t\beta$ band (Fig.\, S5), but both also undergo a similar SDW and/or charge-density wave (CDW) transition around 100-150~K, as observed in transport measurements \cite{Chen2023,li2017}. The mechanism driving this transition has yet to be fully understood, but it has been reported that both bulk \cite{chen2024} and thin-film \cite{gupta2024,Ren2023} LNO327 exhibit an antiferromagnetic SDW with $Q_{\text{SDW}}$ $\sim \pm(0.5,0)$ -- recently established to be slightly incommensurate \cite{gupta2024} -- along the same direction as the Q$_{t\beta}$ translation vector of the $t\beta$ band observed here. A SDW transition along the same direction, with an incommensurate $Q_{\text{SDW}}$ $= \pm(0.62,0)$, has also been reported earlier for LNO4310 and suggested as FS-driven, albeit based on different DFT results \cite{zhang2020}. These values of $Q_{\text{SDW}}$ from bulk sensitive measurements such as RXS \cite{chen2024,gupta2024,Ren2023} and neutron scattering \cite{zhang2020} match very well the range of $Q_{t\beta}$ measured by ARPES [Fig.\,\ref{Fig3}(c)], and are also consistent with our observation that trilayer nickelates tend to have a larger volume and hence a larger $Q$ vector. Such agreement between $Q_{\text{SDW}}$ and $Q_{t\beta}$ is an extremely compelling argument in favor of the $t\beta$ band originating from the translational symmetry breaking SDW, via the ZRS-like states.

\begin{figure}[t!]
\includegraphics[width=17cm]{Figures/fig5_v2.png}
\caption{{\bf Doping- and orbital-dependent Fermi surface gapping due to SDW.} \textbf{(a)} Schematics of a NiO$_2$ bilayer with a (0.5,0) SDW order, with the magnetic moments residing on plaquettes of oxygen sites (blue and red color represent opposite spin orientations), and antiferromagnetic coupling between adjacent planes. \textbf{(b)} Experimental FS measured with LV + LH polarized light at 45 eV (left), and 100 eV (middle and right); the Fermi volumes $n_e$ are obtained from the tight-binding fits shown as white lines. \textbf{(c)} Band dispersion as measured by ARPES along the momentum cuts indicated by the thick orange lines in (b),  with the MDC at $E_{\text{F}}$ shown at the top (integrated over a 20~meV window around $E_{\text{F}}$). \textbf{(d)} Simulation of the electronic dispersion along the same cuts as in (c), obtained with using the SDW model described in the main text and S.I. for net magnetic moments per oxygen plaquette $m_p$ = 0.6 (left), 0.4 (middle), and 0.2 (right) $\mu_B$, as correspondingly indicated.}
\label{Fig4}
\end{figure}

\subsection{Doping- and orbital-dependent SDW gap} 
  
 Extrapolating the $t\beta$ FS to the region where it would overlap with the $\alpha$ FS [dashed line in Fig.\,\ref{Fig3}(a) inset], we also observe the formation of a spectral gap, most prominently visible on dispersions slightly off the high-symmetry direction, as shown in Fig.\,\ref{Fig4}(b,c). The existence of similar gaps have been reported previously for both LNO4310 and LNO327 \cite{du2024,Li_2024}, where temperature-dependent studies suggest a relation to SDW/charge-density wave (CDW), again highlighting the universality among multi-layered nickelates. It is important to note that while the intensity of the \( t\beta \) band is heavily modulated by photon energy, the gap opening is not. This is evident in our 45~eV [Fig.\,\ref{Fig4}(b) left] versus 100~eV [Fig.\,\ref{Fig4}(b) middle] FS maps: whereas the \( t\beta \) FS appears only in the 45~eV data, the $\alpha$ pocket exhibits a gap at both photon energies. In contrast, doping significantly affects both the prominence of the \( t\beta \) band and the gap size: higher electron counting leads to a larger gap and a more pronounced \( t\beta \) pocket, while increased hole doping closes the gap and reduces the coherence of the whole \( t\beta \) band (Fig.\,S12). In this section, with the help of a SDW bilayer model, we show that both the gap opening and the \( t\beta \) band are features stemming from a SDW FS reconstruction driven by nesting specifically of the ZRS-like states. 

We expand our model to a SDW supercell containing eight Ni atoms per plane, with the chemical potential chosen such that the $\alpha$ and $\beta$ pockets overlap, allowing for a $(0.5,0)$ commensurate nesting for simplicity (SI Sect.\,S12). 
From our autocorrelation result in Fig.\,\ref{Fig3}(b), we established that it is the ZRS-like states that are most strongly scattered by the SDW. Hence, to model the SDW we focus on the oxygen sublattice and 
impose a small net magnetic moment on the group of four O \(p_x/p_y\) orbitals surrounding each Ni in a ZRS-like combination, with an overall \mbox{`~up~-~0~-~down~-~0~'} arrangement of the oxygen-plaquette net magnetic moments [where ‘0’ refers to a disordered spin, see Fig.\,\ref{Fig4}(a)]; this is consistent with the magnetic order centered at the Ni atoms as reported from x-ray scattering \cite{gupta2024}. 

Since the wavefunctions of the $\alpha$ and $\beta$ FS are orthogonal, they cannot hybridize even when they overlap. For nesting to lead to a gap opening, the SDW is thus required to have a different phase on adjacent layers within our minimal model, i.e. with the layers being antiferromagnetically coupled as depicted in Fig.\,\ref{Fig4}(a), and consistent with scattering results~\cite{gupta2024,Ren2023,chen2024}. By including this out-of-phase antiferromagnetic coupling in our bilayer model, the $\beta$ FS is translated to reproduce the $t\beta$ FS [cut along green line in Fig.\,S13(c)], and a gap opens near $E_{\text{F}}$ where the $\alpha$ and the $t\beta$ FS overlap [Fig.\,\ref{Fig4}(d)]. We note that since ARPES measurements average over orthogonal magnetic domains of roughly 300~\AA\ \cite{gupta2024}, the gap is most noticeable slightly away from the high-symmetry directions, with the gap size increasing with the net magnetic moment per oxygen plaquette $m_p\!=\![0.2,0.6]~\mu_B$, consistent with the observed ordered magnetic moment of around 0.5~$\mu_B$ in $\mu$SR experiments \cite{khasanov2025,Chen2024_PRL}.

Finally, we should also emphasize that despite having used a commensurate SDW bilayer model for tractability, the key ingredients for enabling the observed FS reconstruction are (i) an in-plane translational symmetry breaking that folds the $\beta$ pocket onto the $\alpha$ pocket, as well as (ii) an out-of-plane mirror symmetry breaking that permits hybridization between $\alpha$ and $\beta$~bands -- regardless of commensurability. More importantly, the strong doping-dependent experimental SDW signatures, namely the $t\beta$ FS and the spectral gap, establish a universal connection between magnetism and the low-energy electronic structure across this broad family of superconducting nickelates.

\subsection{Discussion} 

In this work we have demonstrated that, within a certain range of dopings, the key features of the low-energy electronic structure of the layered nickelates are in fact universal among both the 2222-LNO327 and 1313-LNO327 polymorphs, as well as the related compound LNO4310. This includes the shape and number of bands crossing $E_{\text{F}}$, their momentum-dependent orbital character, and the density-wave order induced FS reconstruction. This universality is unexpected in light of the stark differences in local crystallographic environments realised among the various layered structures, and may be explained by conjecturing a \textit{layer-dependent Mott transition} on the NiO$_2$ monolayer and the inner layer of the trilayer, such that the remaining low-energy electronic states can be captured in an effective bilayer model.  Independent of the degree of correlation in the system, the quasiparticles at $E_{\text{F}}$ will reflect both Luttinger's theorem and the underlying crystallographic symmetry, giving rise to agreement with the single-particle FS. In contrast, the differences in electronic structure among the various multilayer nickelates investigated here become apparent only away from $E_{\text{F}}$, for example in their valence band spectra [see Fig.\,\ref{Fig1}(c) and Fig.\,S10]. 

From our study of ARPES linear dichroism we have demonstrated the importance of the in-plane oxygen orbitals in the description of the first electron-removal state of the nickelates. This is consistent with theoretical descriptions of the cuprates, in which the low-energy electronic structure is dominated by ZRS or, more generally, bond-centered (i.e., oxygen centered) 3SP~\cite{Emery1988}. In cuprates, ZRS and 3SP are equivalent at $(\pi/2,\pi/2)$, but the latter provides a better description in other parts of the BZ~\cite{Lau2011}. This is particularly evident in the present case of LNO327, where our ARPES dichroism results directly reveal the evolution of the first electron-removal states from $d_{x^2-y^2}$ symmetry, along the direction at 45$^{\circ}$ with respect to the Ni-O-Ni bonds, to $d_{3x^2-r^2}$ and $d_{3y^2-r^2}$ symmetry along the Ni-O-Ni bond directions (Fig.\,\ref{Fig3}). Our ARPES measurements also indicate that it is precisely these former electronic states which are the most strongly scattered by SDW order, resulting in a gapping of the $\alpha$ pocket (Fig.\,\ref{Fig4}), and in the $t\beta$ FS fragments observed in the whole family of multilayer nickelate superconductors (Fig.\,\ref{Fig2}). The remarkable agreement between our ARPES measurements and the simple SDW tight-binding bilayer model with magnetic moments residing on oxygen sites (Fig.\,\ref{Fig4} and Fig.\,S13) suggests that the charge carriers that mediate SDW, and potentially also superconductivity, are distributed through a network of ZRS-like oxygen orbitals. In this framework the doping dependence of the SDW and associated FS reconstruction -- which both weaken upon hole doping -- can be understood as the result of an increase in the oxygen content. Recent ARPES studies have shown that ambient-pressure superconducting LNO327 thin films exhibit a hole-doped character relative to their bulk counterparts ~\cite{Li2025a,shen2025,wang2025}. Additionally, oxygen annealing has been found to be essential for inducing superconductivity in both thin films and bulk LNO327~\cite{Li2025,shen2025,wang2025,Puphal2024,Ko2024,zhou2024}, and theoretical studies indicate that oxygen vacancies have a significant impact on both the electronic structure and pairing strength~\cite{sui2024,lu2025,Wang_2025}. All together, these results suggest that the ZRS population plays a crucial role in determining whether the ground state favors either SDW or superconductivity, with the latter emerging upon hole doping, while the SDW -- which is generally considered to compete with superconductivity -- is suppressed.

The degree to which high-temperature superconductivity in the cuprates and nickelates is comparable -- particularly regarding pairing mechanism and symmetry -- is not at all obvious given the multiorbital nature of the nickelates, where both Ni $d_{x^2-y^2}$ and  $d_{3z^2-r^2}$  can play a role in the low-energy electronic structure.  In broad agreement with recent proposals~\cite{qin2024}, our results establish that the underlying physics of the nickelates still stems from ZRS-cuprate-like states in which hybridization with the planar oxygens plays a central role. Within this picture, oxygen content may control superconductivity by modulating the competing charge or spin order correlations.


\bibliographystyle{naturemag}
\bibliography{LNO327}

\begin{thebibliography}{10}
\expandafter\ifx\csname url\endcsname\relax
  \def\url#1{\texttt{#1}}\fi
\expandafter\ifx\csname urlprefix\endcsname\relax\def\urlprefix{URL }\fi
\providecommand{\bibinfo}[2]{#2}
\providecommand{\eprint}[2][]{\url{#2}}

\bibitem{Bednorz1986}
\bibinfo{author}{Bednorz, J.~G.} \& \bibinfo{author}{M{\"u}ller, K.~A.}
\newblock \bibinfo{title}{{Possible high ${T}_{c}$ superconductivity in the {B}a--{L}a--{C}u--{O} system}}.
\newblock \emph{\bibinfo{journal}{Z. Phys. B Con. Mat.}} \textbf{\bibinfo{volume}{64}}, \bibinfo{pages}{189--193} (\bibinfo{year}{1986}).

\bibitem{Maeno1994}
\bibinfo{author}{Maeno, Y.} \emph{et~al.}
\newblock \bibinfo{title}{Superconductivity in a layered perovskite without copper}.
\newblock \emph{\bibinfo{journal}{Nature}} \textbf{\bibinfo{volume}{372}}, \bibinfo{pages}{532--534} (\bibinfo{year}{1994}).

\bibitem{Yan2015}
\bibinfo{author}{Yan, Y.} \emph{et~al.}
\newblock \bibinfo{title}{{Electron-Doped Sr$_2$IrO$_4$: An Analogue of Hole-Doped Cuprate Superconductors Demonstrated by Scanning Tunneling Microscopy}}.
\newblock \emph{\bibinfo{journal}{Phys. Rev. X}} \textbf{\bibinfo{volume}{5}}, \bibinfo{pages}{041018} (\bibinfo{year}{2015}).

\bibitem{Kim2015}
\bibinfo{author}{Kim, Y.~K.}, \bibinfo{author}{Sung, N.~H.}, \bibinfo{author}{Denlinger, J.~D.} \& \bibinfo{author}{Kim, B.~J.}
\newblock \bibinfo{title}{{Observation of a $d$-wave gap in electron-doped Sr$_2$IrO$_4$}}.
\newblock \emph{\bibinfo{journal}{Nat. Phys.}} \textbf{\bibinfo{volume}{12}}, \bibinfo{pages}{37--41} (\bibinfo{year}{2015}).

\bibitem{Anisimov1999}
\bibinfo{author}{Anisimov, V.~I.}, \bibinfo{author}{Bukhvalov, D.} \& \bibinfo{author}{Rice, T.~M.}
\newblock \bibinfo{title}{Electronic structure of possible nickelate analogs to the cuprates}.
\newblock \emph{\bibinfo{journal}{Phys. Rev. B}} \textbf{\bibinfo{volume}{59}}, \bibinfo{pages}{7901--7906} (\bibinfo{year}{1999}).

\bibitem{Chaloupka2008}
\bibinfo{author}{Chaloupka, J.} \& \bibinfo{author}{Khaliullin, G.}
\newblock \bibinfo{title}{{Orbital Order and Possible Superconductivity in LaNiO$_3$/La$M$O$_3$ Superlattices}}.
\newblock \emph{\bibinfo{journal}{Phys. Rev. Lett.}} \textbf{\bibinfo{volume}{100}}, \bibinfo{pages}{016404} (\bibinfo{year}{2008}).

\bibitem{Hansmann2009}
\bibinfo{author}{Hansmann, P.} \emph{et~al.}
\newblock \bibinfo{title}{{Turning a Nickelate Fermi Surface into a Cuprate-like One through Heterostructuring}}.
\newblock \emph{\bibinfo{journal}{Phys. Rev. Lett.}} \textbf{\bibinfo{volume}{103}}, \bibinfo{pages}{016401} (\bibinfo{year}{2009}).

\bibitem{Li2019}
\bibinfo{author}{Li, D.} \emph{et~al.}
\newblock \bibinfo{title}{Superconductivity in an infinite-layer nickelate}.
\newblock \emph{\bibinfo{journal}{Nature}} \textbf{\bibinfo{volume}{572}}, \bibinfo{pages}{624--627} (\bibinfo{year}{2019}).

\bibitem{Sun2023}
\bibinfo{author}{Sun, H.} \emph{et~al.}
\newblock \bibinfo{title}{{Signatures of superconductivity near 80~K in a nickelate under high pressure}}.
\newblock \emph{\bibinfo{journal}{Nature}} \textbf{\bibinfo{volume}{621}}, \bibinfo{pages}{493--498} (\bibinfo{year}{2023}).

\bibitem{Li2025}
\bibinfo{author}{Li, F.} \emph{et~al.}
\newblock \bibinfo{title}{{Ambient pressure growth of bilayer nickelate single crystals with superconductivity over 90 K under high pressure}} (\bibinfo{year}{2025}).
\newblock \eprint{arXiv:2501.14584}.

\bibitem{zhu2024}
\bibinfo{author}{Zhu, Y.}, \bibinfo{author}{Peng, D.} \& \bibinfo{author}{Zhang, E.}
\newblock \bibinfo{title}{{Superconductivity in pressurized trilayer $\mathrm{La}_{4}\mathrm{N}{\mathrm{i}}_{3}{\mathrm{O}}_{10-\delta}$ single crystals}}.
\newblock \emph{\bibinfo{journal}{Nature}} \textbf{\bibinfo{volume}{631}}, \bibinfo{pages}{531–536} (\bibinfo{year}{2024}).

\bibitem{Ko2024}
\bibinfo{author}{Ko, E.~K.} \emph{et~al.}
\newblock \bibinfo{title}{{Signatures of ambient pressure superconductivity in thin film La$_3$Ni$_2$O$_7$}}.
\newblock \emph{\bibinfo{journal}{Nature}} \textbf{\bibinfo{volume}{638}}, \bibinfo{pages}{935--940} (\bibinfo{year}{2025}).

\bibitem{zhou2024}
\bibinfo{author}{Zhou, G.}, \bibinfo{author}{Lv, W.}, \bibinfo{author}{Wang, H.} \emph{et~al.}
\newblock \bibinfo{title}{{Ambient-pressure superconductivity onset above 40 K in (La,Pr)$_3$Ni$_2$O$_7$ films}}.
\newblock \emph{\bibinfo{journal}{Nature}} \textbf{\bibinfo{volume}{640}}, \bibinfo{pages}{641--646} (\bibinfo{year}{2025}).

\bibitem{Chen2023}
\bibinfo{author}{Chen, X.} \emph{et~al.}
\newblock \bibinfo{title}{{Polymorphism in the Ruddlesden--Popper Nickelate La$_3$Ni$_2$O$_7$: Discovery of a Hidden Phase with Distinctive Layer Stacking}}.
\newblock \emph{\bibinfo{journal}{J. Am. Chem. Soc.}} \textbf{\bibinfo{volume}{146}}, \bibinfo{pages}{3640--3645} (\bibinfo{year}{2024}).

\bibitem{Puphal2024}
\bibinfo{author}{Puphal, P.} \emph{et~al.}
\newblock \bibinfo{title}{{Unconventional Crystal Structure of the High-Pressure Superconductor ${\mathrm{La}}_{3}{\mathrm{Ni}}_{2}{\mathrm{O}}_{7}$}}.
\newblock \emph{\bibinfo{journal}{Phys. Rev. Lett.}} \textbf{\bibinfo{volume}{133}}, \bibinfo{pages}{146002} (\bibinfo{year}{2024}).

\bibitem{Yang2023}
\bibinfo{author}{Yang, J.} \emph{et~al.}
\newblock \bibinfo{title}{{Orbital-dependent electron correlation in double-layer nickelate La$_3$Ni$_2$O$_7$}}.
\newblock \emph{\bibinfo{journal}{Nat. Commun.}} \textbf{\bibinfo{volume}{15}}, \bibinfo{pages}{4373} (\bibinfo{year}{2024}).

\bibitem{Abadi2024}
\bibinfo{author}{Abadi, S.} \emph{et~al.}
\newblock \bibinfo{title}{{Electronic Structure of the Alternating Monolayer-Trilayer Phase of ${\mathrm{La}}_{3}{\text{Ni}}_{2}{\mathrm{O}}_{7}$}}.
\newblock \emph{\bibinfo{journal}{Phys. Rev. Lett.}} \textbf{\bibinfo{volume}{134}}, \bibinfo{pages}{126001} (\bibinfo{year}{2025}).

\bibitem{du2024}
\bibinfo{author}{Du, X.} \emph{et~al.}
\newblock \bibinfo{title}{{Correlated Electronic Structure and Density-Wave Gap in Trilayer Nickelate La$_4$Ni$_3$O$_{10}$}} (\bibinfo{year}{2024}).
\newblock \eprint{arXiv:2405.19853}.

\bibitem{li2017}
\bibinfo{author}{Li, H.} \emph{et~al.}
\newblock \bibinfo{title}{{Fermiology and electron dynamics of trilayer nickelate La$_4$Ni$_3$O$_{10}$}}.
\newblock \emph{\bibinfo{journal}{Nature Communications}} \textbf{\bibinfo{volume}{8}}, \bibinfo{pages}{704} (\bibinfo{year}{2017}).

\bibitem{Zhang2024}
\bibinfo{author}{Zhang, Y.}, \bibinfo{author}{Lin, L.-F.}, \bibinfo{author}{Moreo, A.}, \bibinfo{author}{Maier, T.~A.} \& \bibinfo{author}{Dagotto, E.}
\newblock \bibinfo{title}{{Electronic structure, self-doping, and superconducting instability in the alternating single-layer trilayer stacking nickelates ${\mathrm{La}}_{3}{\mathrm{Ni}}_{2}{\mathrm{O}}_{7}$}}.
\newblock \emph{\bibinfo{journal}{Phys. Rev. B}} \textbf{\bibinfo{volume}{110}}, \bibinfo{pages}{L060510} (\bibinfo{year}{2024}).

\bibitem{LaBollita2024}
\bibinfo{author}{LaBollita, H.}, \bibinfo{author}{Bag, S.}, \bibinfo{author}{Kapeghian, J.} \& \bibinfo{author}{Botana, A.~S.}
\newblock \bibinfo{title}{{Electronic correlations, layer distinction, and electron doping in the alternating single-layer--trilayer ${\mathrm{La}}_{3}{\mathrm{Ni}}_{2}{\mathrm{O}}_{7}$ polymorph}}.
\newblock \emph{\bibinfo{journal}{Phys. Rev. B}} \textbf{\bibinfo{volume}{110}}, \bibinfo{pages}{155145} (\bibinfo{year}{2024}).

\bibitem{Li_2024}
\bibinfo{author}{Li, Y.} \emph{et~al.}
\newblock \bibinfo{title}{{Electronic Correlation and Pseudogap-Like Behavior of High-Temperature Superconductor La$_3$Ni$_2$O$_7$}}.
\newblock \emph{\bibinfo{journal}{Chinese Physics Letters}} \textbf{\bibinfo{volume}{41}}, \bibinfo{pages}{087402} (\bibinfo{year}{2024}).

\bibitem{Fournier2010}
\bibinfo{author}{Fournier, D.} \emph{et~al.}
\newblock \bibinfo{title}{{Loss of nodal quasiparticle integrity in underdoped YBa$_2$Cu$_3$O$_{6+x}$}}.
\newblock \emph{\bibinfo{journal}{Nat. Phys.}} \textbf{\bibinfo{volume}{6}}, \bibinfo{pages}{905--911} (\bibinfo{year}{2010}).

\bibitem{Ding1996}
\bibinfo{author}{Ding, H.} \emph{et~al.}
\newblock \bibinfo{title}{{Electronic Excitations in ${\mathrm{Bi}}_{2}{\mathrm{Sr}}_{2}\mathrm{Ca}{\mathrm{Cu}}_{2}{O}_{8}$: Fermi Surface, Dispersion, and Absence of Bilayer Splitting}}.
\newblock \emph{\bibinfo{journal}{Phys. Rev. Lett.}} \textbf{\bibinfo{volume}{76}}, \bibinfo{pages}{1533--1536} (\bibinfo{year}{1996}).

\bibitem{Feng2001}
\bibinfo{author}{Feng, D.~L.} \emph{et~al.}
\newblock \bibinfo{title}{{Bilayer Splitting in the Electronic Structure of Heavily Overdoped ${\mathrm{Bi}}_{2}{\mathrm{Sr}}_{2}{\mathrm{CaCu}}_{2}{O}_{8+\ensuremath{\delta}}$}}.
\newblock \emph{\bibinfo{journal}{Phys. Rev. Lett.}} \textbf{\bibinfo{volume}{86}}, \bibinfo{pages}{5550--5553} (\bibinfo{year}{2001}).

\bibitem{Chuang2001}
\bibinfo{author}{Chuang, Y.-D.} \emph{et~al.}
\newblock \bibinfo{title}{{Doubling of the Bands in Overdoped ${{\mathrm{Bi}}_{2}{\mathrm{Sr}}_{2}{\mathrm{CaCu}}_{2}O}_{8+\mathit{\ensuremath{\delta}}}$: Evidence for $\mathit{c}$-Axis Bilayer Coupling}}.
\newblock \emph{\bibinfo{journal}{Phys. Rev. Lett.}} \textbf{\bibinfo{volume}{87}}, \bibinfo{pages}{117002} (\bibinfo{year}{2001}).

\bibitem{Luo2023}
\bibinfo{author}{Luo, X.} \emph{et~al.}
\newblock \bibinfo{title}{{Electronic origin of high superconducting critical temperature in trilayer cuprates}}.
\newblock \emph{\bibinfo{journal}{Nat. Phys.}} \textbf{\bibinfo{volume}{19}}, \bibinfo{pages}{1841--1847} (\bibinfo{year}{2023}).

\bibitem{Zhou2025}
\bibinfo{author}{Zhou, Y.} \emph{et~al.}
\newblock \bibinfo{title}{{Investigations of key issues on the reproducibility of high- T$_c$ superconductivity emerging from compressed La$_3$Ni$_2$O$_7$}}.
\newblock \emph{\bibinfo{journal}{Matter Radiat. Extrem.}} \textbf{\bibinfo{volume}{10}} (\bibinfo{year}{2025}).

\bibitem{Wang_2025}
\bibinfo{author}{Wang, Y.}, \bibinfo{author}{Zhang, Y.} \& \bibinfo{author}{Jiang, K.}
\newblock \bibinfo{title}{{Electronic structure and disorder effect of La$_3$Ni$_2$O$_7$ superconductor}}.
\newblock \emph{\bibinfo{journal}{Chinese Physics B}} \textbf{\bibinfo{volume}{34}}, \bibinfo{pages}{047105} (\bibinfo{year}{2025}).

\bibitem{Day2019}
\bibinfo{author}{Day, R.~P.}, \bibinfo{author}{Zwartsenberg, B.}, \bibinfo{author}{Elfimov, I.~S.} \& \bibinfo{author}{Damascelli, A.}
\newblock \bibinfo{title}{{Computational framework chinook for angle-resolved photoemission spectroscopy}}.
\newblock \emph{\bibinfo{journal}{npj Quantum Mater.}} \textbf{\bibinfo{volume}{4}}, \bibinfo{pages}{54} (\bibinfo{year}{2019}).

\bibitem{Lechermann2024}
\bibinfo{author}{Lechermann, F.}, \bibinfo{author}{B\"otzel, S.} \& \bibinfo{author}{Eremin, I.~M.}
\newblock \bibinfo{title}{{Electronic instability, layer selectivity, and Fermi arcs in ${\text{La}}_{3}{\text{Ni}}_{2}{\text{O}}_{7}$}}.
\newblock \emph{\bibinfo{journal}{Phys. Rev. Mater.}} \textbf{\bibinfo{volume}{8}}, \bibinfo{pages}{074802} (\bibinfo{year}{2024}).

\bibitem{Zaanen1985}
\bibinfo{author}{Zaanen, J.}, \bibinfo{author}{Sawatzky, G.~A.} \& \bibinfo{author}{Allen, J.~W.}
\newblock \bibinfo{title}{Band gaps and electronic structure of transition-metal compounds}.
\newblock \emph{\bibinfo{journal}{Phys. Rev. Lett.}} \textbf{\bibinfo{volume}{55}}, \bibinfo{pages}{418--421} (\bibinfo{year}{1985}).

\bibitem{Johnston2014}
\bibinfo{author}{Johnston, S.}, \bibinfo{author}{Mukherjee, A.}, \bibinfo{author}{Elfimov, I.}, \bibinfo{author}{Berciu, M.} \& \bibinfo{author}{Sawatzky, G.~A.}
\newblock \bibinfo{title}{Charge disproportionation without charge transfer in the rare-earth-element nickelates as a possible mechanism for the metal-insulator transition}.
\newblock \emph{\bibinfo{journal}{Phys. Rev. Lett.}} \textbf{\bibinfo{volume}{112}}, \bibinfo{pages}{106404} (\bibinfo{year}{2014}).

\bibitem{Zhang1988}
\bibinfo{author}{Zhang, F.~C.} \& \bibinfo{author}{Rice, T.~M.}
\newblock \bibinfo{title}{{Effective Hamiltonian for the superconducting Cu oxides}}.
\newblock \emph{\bibinfo{journal}{Phys. Rev. B}} \textbf{\bibinfo{volume}{37}}, \bibinfo{pages}{3759--3761} (\bibinfo{year}{1988}).

\bibitem{Lau2011}
\bibinfo{author}{Lau, B.}, \bibinfo{author}{Berciu, M.} \& \bibinfo{author}{Sawatzky, G.~A.}
\newblock \bibinfo{title}{{High-Spin Polaron in Lightly Doped ${\mathrm{CuO}}_{2}$ Planes}}.
\newblock \emph{\bibinfo{journal}{Phys. Rev. Lett.}} \textbf{\bibinfo{volume}{106}}, \bibinfo{pages}{036401} (\bibinfo{year}{2011}).

\bibitem{Emery1988}
\bibinfo{author}{Emery, V.~J.} \& \bibinfo{author}{Reiter, G.}
\newblock \bibinfo{title}{Mechanism for high-temperature superconductivity}.
\newblock \emph{\bibinfo{journal}{Phys. Rev. B}} \textbf{\bibinfo{volume}{38}}, \bibinfo{pages}{4547--4556} (\bibinfo{year}{1988}).

\bibitem{Glazer1972}
\bibinfo{author}{Glazer, A.~M.}
\newblock \bibinfo{title}{The classification of tilted octahedra in perovskites}.
\newblock \emph{\bibinfo{journal}{Acta Crystallographica Section B}} \textbf{\bibinfo{volume}{28}}, \bibinfo{pages}{3384--3392} (\bibinfo{year}{1972}).

\bibitem{chen2024}
\bibinfo{author}{Chen, X.} \emph{et~al.}
\newblock \bibinfo{title}{{Electronic and magnetic excitations in La$_3$Ni$_2$O$_7$}}.
\newblock \emph{\bibinfo{journal}{Nature Communications}} \textbf{\bibinfo{volume}{15}}, \bibinfo{pages}{9597} (\bibinfo{year}{2024}).

\bibitem{gupta2024}
\bibinfo{author}{Gupta, N.~K.}, \bibinfo{author}{Gong, R.}, \bibinfo{author}{Wu, Y.} \emph{et~al.}
\newblock \bibinfo{title}{{Anisotropic spin stripe domains in bilayer La$_3$Ni$_2$O$_7$}}.
\newblock \emph{\bibinfo{journal}{Nature Communications}} \textbf{\bibinfo{volume}{16}}, \bibinfo{pages}{6560} (\bibinfo{year}{2025}).

\bibitem{Ren2023}
\bibinfo{author}{Ren, X.} \emph{et~al.}
\newblock \bibinfo{title}{{Resolving the electronic ground state of La$_3$Ni$_2$O$_{7-\delta}$ films}}.
\newblock \emph{\bibinfo{journal}{Commun. Phys.}} \textbf{\bibinfo{volume}{8}}, \bibinfo{pages}{52} (\bibinfo{year}{2025}).

\bibitem{zhang2020}
\bibinfo{author}{Zhang, J.} \emph{et~al.}
\newblock \bibinfo{title}{Intertwined density waves in a metallic nickelate}.
\newblock \emph{\bibinfo{journal}{Nature Communications}} \textbf{\bibinfo{volume}{11}}, \bibinfo{pages}{6003} (\bibinfo{year}{2020}).

\bibitem{khasanov2025}
\bibinfo{author}{Khasanov, R.} \emph{et~al.}
\newblock \bibinfo{title}{{Pressure-enhanced splitting of density wave transitions in La$_3$Ni$_2$O$_{7–\delta}$}}.
\newblock \emph{\bibinfo{journal}{Nature Physics}} \textbf{\bibinfo{volume}{21}}, \bibinfo{pages}{430--436} (\bibinfo{year}{2025}).

\bibitem{Chen2024_PRL}
\bibinfo{author}{Chen, K.} \emph{et~al.}
\newblock \bibinfo{title}{{Evidence of Spin Density Waves in ${\mathrm{La}}_{3}{\mathrm{Ni}}_{2}{\mathrm{O}}_{7\ensuremath{-}\ensuremath{\delta}}$}}.
\newblock \emph{\bibinfo{journal}{Phys. Rev. Lett.}} \textbf{\bibinfo{volume}{132}}, \bibinfo{pages}{256503} (\bibinfo{year}{2024}).

\bibitem{Li2025a}
\bibinfo{author}{Li, P.} \emph{et~al.}
\newblock \bibinfo{title}{{Angle-resolved photoemission spectroscopy of superconducting (La,Pr)$_3$Ni$_2$O$_7$/SrLaAlO$_4$ heterostructures}}.
\newblock \emph{\bibinfo{journal}{National Science Review}}  (\bibinfo{year}{2025}).

\bibitem{shen2025}
\bibinfo{author}{Shen, J.} \emph{et~al.}
\newblock \bibinfo{title}{{Nodeless superconducting gap and electron-boson coupling in (La,Pr,Sm)$_{3}$Ni$_2$O$_7$ films}} (\bibinfo{year}{2025}).
\newblock \eprint{arXiv:2502.17831}.

\bibitem{wang2025}
\bibinfo{author}{Wang, B.~Y.} \emph{et~al.}
\newblock \bibinfo{title}{{Electronic structure of compressively strained thin film La$_2$PrNi$_2$O$_7$}} (\bibinfo{year}{2025}).
\newblock \eprint{arXiv:2504.16372}.

\bibitem{sui2024}
\bibinfo{author}{Sui, X.} \emph{et~al.}
\newblock \bibinfo{title}{{Electronic properties of the bilayer nickelates ${R}_{3}\mathrm{N}{\mathrm{i}}_{2}{\mathrm{O}}_{7}$ with oxygen vacancies ($R=\mathrm{La}$ or Ce)}}.
\newblock \emph{\bibinfo{journal}{Phys. Rev. B}} \textbf{\bibinfo{volume}{109}}, \bibinfo{pages}{205156} (\bibinfo{year}{2024}).

\bibitem{lu2025}
\bibinfo{author}{Lu, C.}, \bibinfo{author}{Zhang, M.}, \bibinfo{author}{Pan, Z.}, \bibinfo{author}{Wu, C.} \& \bibinfo{author}{Yang, F.}
\newblock \bibinfo{title}{{Impact of Pressure and Apical Oxygen Vacancies on Superconductivity in La$_3$Ni$_2$O$_7$}} (\bibinfo{year}{2025}).
\newblock \eprint{arXiv:2502.14324}.

\bibitem{qin2024}
\bibinfo{author}{Jiang, G.}, \bibinfo{author}{Qin, C.}, \bibinfo{author}{Foyevtsova, K.} \emph{et~al.}
\newblock \bibinfo{title}{{Intertwined charge and spin instability of La$_3$Ni$_2$O$_7$}}.
\newblock \emph{\bibinfo{journal}{Science China Physics, Mechanics \& Astronomy}} \textbf{\bibinfo{volume}{68}}, \bibinfo{pages}{297411} (\bibinfo{year}{2025}).

\bibitem{Zhang22020}
\bibinfo{author}{Zhang, J.} \emph{et~al.}
\newblock \bibinfo{title}{{High oxygen pressure floating zone growth and crystal structure of the metallic nickelates ${R}_{4}{\mathrm{Ni}}_{3}{\mathrm{O}}_{10}$ ($R=\mathrm{La},\mathrm{Pr}$)}}.
\newblock \emph{\bibinfo{journal}{Phys. Rev. Mater.}} \textbf{\bibinfo{volume}{4}}, \bibinfo{pages}{083402} (\bibinfo{year}{2020}).

\bibitem{QE-2009}
\bibinfo{author}{Giannozzi, P.} \emph{et~al.}
\newblock \bibinfo{title}{Quantum espresso: a modular and open-source software project for quantum simulations of materials}.
\newblock \emph{\bibinfo{journal}{Journal of Physics: Condensed Matter}} \textbf{\bibinfo{volume}{21}}, \bibinfo{pages}{395502 (19pp)} (\bibinfo{year}{2009}).

\bibitem{QE-2017}
\bibinfo{author}{Giannozzi, P.} \emph{et~al.}
\newblock \bibinfo{title}{Advanced capabilities for materials modelling with quantum espresso}.
\newblock \emph{\bibinfo{journal}{Journal of Physics: Condensed Matter}} \textbf{\bibinfo{volume}{29}}, \bibinfo{pages}{465901} (\bibinfo{year}{2017}).

\bibitem{PAW-1994}
\bibinfo{author}{Bl\"ochl, P.~E.}
\newblock \bibinfo{title}{Projector augmented-wave method}.
\newblock \emph{\bibinfo{journal}{Phys. Rev. B}} \textbf{\bibinfo{volume}{50}}, \bibinfo{pages}{17953--17979} (\bibinfo{year}{1994}).

\bibitem{PBE-1996}
\bibinfo{author}{Perdew, J.~P.}, \bibinfo{author}{Burke, K.} \& \bibinfo{author}{Ernzerhof, M.}
\newblock \bibinfo{title}{Generalized gradient approximation made simple}.
\newblock \emph{\bibinfo{journal}{Phys. Rev. Lett.}} \textbf{\bibinfo{volume}{77}}, \bibinfo{pages}{3865--3868} (\bibinfo{year}{1996}).

\bibitem{wannier90}
\bibinfo{author}{Mostofi, A.~A.} \emph{et~al.}
\newblock \bibinfo{title}{An updated version of wannier90: A tool for obtaining maximally-localised wannier functions}.
\newblock \emph{\bibinfo{journal}{Computer Physics Communications}} \textbf{\bibinfo{volume}{185}}, \bibinfo{pages}{2309--2310} (\bibinfo{year}{2014}).

\end{thebibliography}
\subsection{Methods}

High-quality single crystals of La$_3$Ni$_2$O$_7$ were grown by the high-pressure floating zone method \cite{Chen2023,Zhang22020}. All measured crystals were preliminarily screened by XRD for a rough estimate of phase composition and purity. ARPES experiments were performed at the Quantum Materials Spectroscopy Centre (QMSC) beamline of the Canadian Light Source (CLS), using a Scienta R4000 hemispherical analyzer with angular and energy resolutions better then 0.1$^o$ and 15 meV, respectively. The selected samples, aligned ex situ by conventional Laue diffraction, were then cleaved and measured in the ARPES chamber at a temperature of 17\,K  and a base pressure lower than 5x$10^{-11}$ torr. ARPES experiments were performed using circular, linear vertical, and linear horizontal polarizations, over a range of photon energies from 45 to 100 eV as indicated in the text. 
\\
DFT calculations were performed with the Quantum Espresso code \cite{QE-2009,QE-2017}, using the projector augmented wave (PAW) method \cite{PAW-1994}. The Perdew-Burke-Ernzerhof (PBE) functional \cite{PBE-1996} was used for the exchange-correlation energy; the kinetic energy cutoff for wavefunctions was set to 100 eV, and the BZ was sampled using a 12x12x4 $k$-grid. The Wannier90 package was used to calculate the initial guess for tight-binding parameters \cite{wannier90}. The two effective bilayer tight-binding models were generated as follows: the first one contains only Ni $e_g$ orbitals, while the second one has additional O $p_x$/$p_y$ orbitals which are $\sigma$-bonded to the Ni $e_g$. 
The SDW tight-binding model is a bilayer system with eight Ni sites per plane, each containing $e_g$ orbitals which are $\sigma$-bonded to $p_x$ and $p_y$ orbitals within the plane, and to $p_z$ orbitals between the planes. The ARPES intensity is then calculated using the \textit{chinook} package \cite{Day2019} (see also SI Sect.\,S6, S7 and S12).

\subsection*{Acknowledgments}

We gratefully acknowledge T.P.~Devereaux, D.G.~Hawthorn and Z.~Zhu for valuable discussions and insights related to their work. This research was undertaken thanks in part to funding from the Max Planck–UBC–UTokyo Centre for Quantum Materials and the Canada First Research Excellence Fund (CFREF), Quantum Materials and Future Technologies. Work at Argonne National Laboratory (crystal growth, x-ray characterization) was supported by the U.S. DOE Office of Science, Basic Energy Sciences, Materials Science and Engineering Division. This project is also funded by the Natural Sciences and Engineering Research Council of Canada (NSERC), the Canada Foundation for Innovation (CFI); the Department of National Defence (DND); the British Columbia Knowledge Development Fund (BCKDF); the Mitacs Accelerate Program; the QuantEmX Program of the Institute for Complex Adaptive Matter (ICAM); the Moore EPiQS Program (A.D.); the Canada Research Chairs (CRC) Program (A.D.); and the CIFAR Quantum Materials Program (A.D.). Use of the Canadian Light Source (Quantum Materials Spectroscopy Centre), a national research facility of the University of Saskatchewan, is supported by  CFI, NSERC, the National Research Council (NRC), the Canadian Institutes of Health Research (CIHR), the Government of Saskatchewan and the University of Saskatchewan. S.S. acknowledges the support of the Netherlands Organization for Scientific Research (NWO 019.223EN.014, Rubicon 2022-3).

\end{document}